\documentclass[aps,prl,showpacs,twocolumn,superscriptaddress]{revtex4}
\usepackage{graphicx}
\usepackage{amssymb}
\usepackage{epsfig}
\usepackage{graphicx}
\usepackage{amsmath}
\usepackage{array,color}

\usepackage{soul, xcolor}

\begin{document}

\setstcolor{red} 

\title{Emergence of Quantum Nonmagnetic Insulating Phase in Spin-Orbit Coupled Square Lattices}
\author{Xin Zhang}
\affiliation{Center for Theoretical Physics, Department of Physics,
Capital Normal University, Beijing 100048, China}
\author{Wei Wu}
\affiliation{D\'{e}partement de Physique and RQMP, Universit\'{e} de
Sherbrooke, Sherbrooke, Qu\'{e}bec, Canada}
\author{Gang Li}
\affiliation{Lehrstuhl fuer Theoretische Physick, Universit\"{a}t
Wuerzburg, 97074 Wuerzburg, Germany }
\author{Lin Wen}
\affiliation{College of Physics and Electronic Engineering,
Chongqing Normal University, Chongqing, 401331, China}
\author{Qing Sun}
\affiliation{Center for Theoretical Physics, Department of Physics,
Capital Normal University, Beijing 100048, China}
\author{An-Chun Ji}
\affiliation{Center for Theoretical
Physics, Department of Physics, Capital Normal University, Beijing
100048, China}
\date{{\small \today}}

\begin{abstract}
We investigate the  metal-insulator transition (MIT)  and phase
diagram of the half-filled Fermi Hubbard model  with Rashba-type
spin-orbit coupling (SOC) on a square optical lattice. The interplay
between  the atomic interactions and SOC results in distinctive
features of the MIT. Significantly,  in addition to the diverse spin
ordered phases, a nonmagnetic insulating phase emerges in a
considerably large regime of parameters near the Mott transition.
This phase has a finite single-particle gap but vanishing
magnetization and spin correlation exhibits a power-law scaling,
suggesting a potential algebraic spin-liquid ground state. These
results are confirmed by the non-perturbative cluster dynamical
mean-field theory.
\end{abstract}






\maketitle

The study of quantum many-body effects and new exotic states of
matter are currently amongst the main topics in condensed-matter
physics \cite{Wen,Sachdev}.  During the last few years, the
successful manipulation of  ultracold atoms in optical lattices
~\cite{Greiner,Kohl,Spielman,Jordens,Schneider,Esslinger} and the
experimental progress in the spin-orbit coupling (SOC) of degenerate
atomic gases \cite{Lin,ZhangJY,WangPJ,Cheuk,Qu} have made it
possible to explore diverse quantum phases
\cite{Jaksch1,Lewenstein,Bloch,Goldman11,Zhai,Galitski,Zhou1,Zhai1}.
More recently, optical lattices combined with SOC have attracted
enormous interests. It was shown that SOC plays prominent roles in
many fascinating phenomena, such as non-Abelian interferometry
\cite{Osterloh} and magnetic monopole \cite{Ruseckas,Pietila},
topological phase transitions \cite{Bermudez,Bermudez1,Goldman3},
non-Abelian localization \cite{Satija}, or emerging relativistic
fermions \cite{Goldman1}.

When further competing with strong atomic interactions,  SOC
introduces additional degrees of quantum fluctuation, giving rise to
remarkable many-body ground states. For example, the study of the
superfluid to Mott insulator transition in the Bose-Hubbard model
with synthetic SOC has demonstrated that, Rashba-type SOC can induce
intriguing magnetism in the deep Mott regime
\cite{Grab,Cole,Radic,Cai,Gong,Zhang1,Qian,HLiang,Hickey}, as well
as an exotic superfluid phase with magnetic textures  near  the Mott
transition \cite{Cole,HLiang}. Despite this, the essential
properties of the metal-insulator transition (MIT) of interacting
fermion systems have been less achieved.
\begin{figure}[t]
\begin{center}
\includegraphics[width=0.48\textwidth]{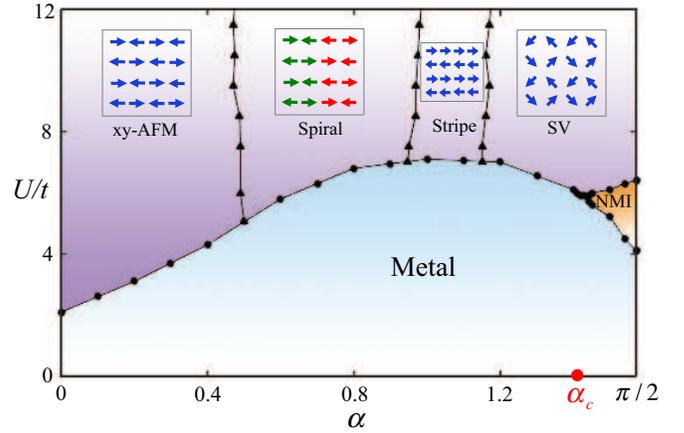}
\end{center}
\caption{(color online). Phase diagram of the half-filled Fermi
Hubbard model  with Rashba-type SOC obtained by the cluster
dynamical mean-field theory with a $2\times2$ cluster at $T=0.05t$.
The solid line with dots is the phase boundary of the MIT. The
purple-colored regions denote the diverse spin ordered phases of
$xy$-antiferromagnet ($xy$-AFM), spiral (the green and red arrows
indicate the spins have up or down $z$-components), stripe, and spin
vortex (SV) in the Mott insulating regime. For $\alpha>\alpha_c$,
there exhibits a nonmagnetic insulating (NMI) phase in the vicinity
of the MIT.
 \label{phase1}}
\end{figure}

In this Letter, we show that SOC can stabilize a quantum nonmagnetic
insulating (NMI) phase in a strongly correlated fermion system. Such
a system described by the spin-orbit (SO) coupled  Fermi Hubbard
model (see Eq. (\ref{Hamiltonian})) has strong implications for
realistic electronic materials \cite{Banerjee}. Our main results are
summarized in Fig. \ref{phase1}, which displays a rich phase
diagram. First, the SOC tends to destroy the conventional
antiferromagnetic fluctuations. This results in the distinctive
features  of the MIT with diverse spin ordered phases occurring on
the side of Mott insulator. Significantly, a NMI phase emerges in
the vicinity of the Mott transition for $\alpha>\alpha_c$ ($\alpha$
is the strength of SOC). This phase possesses a finite single
particle energy gap with vanishing magnetic orders and the spin
correlation function exhibits a power-law scaling, suggesting a
potential algebraic spin-liquid ground state. Recently, enormous
attentions have been paid to the search for the quantum disordered
phase in the interacting fermion systems \cite{Meng,Lu,Clark,Yang}.
Our results formulate a promising new route to achieve this
intriguing quantum state through the SO coupled fermions.
\begin{figure}[t]
\includegraphics[width=0.43\textwidth]{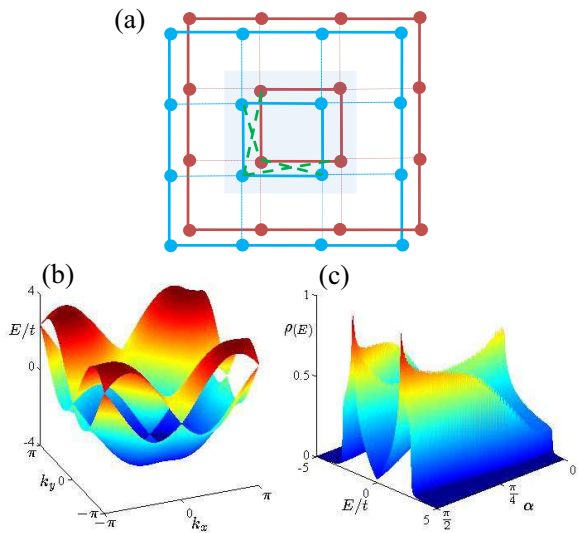}
\caption{(color online). (a) Illustration of the SO coupled square
lattices, which are mapped onto two sets of sublattices for spin up
(red) and down (blue) respectively. The central shaded box denotes
the $2\times2$ cluster, where the dashed lines represent the
spin-flipped hoppings. (b) Single-particle energy spectra for
$\alpha=0.3\pi$. (c) Density of states $\rho(E)$ for non-interacting
fermions with the strength of SOC $\alpha\in[0,\pi/2]$. }
\label{fig2}
\end{figure}

{\it The model.}--The Hamiltonian of a  two-component Fermi gas
subject to an optical square lattice is given by
\begin{eqnarray}
\hat{H}&=&-t\sum_{<ij>}\sum_{\sigma\sigma^\prime}
(\hat{c}^\dag_{i\sigma}\mathcal{R}_{ij}\hat{c}_{j\sigma^\prime}
+{\rm H.c.})\nonumber\\
&&+U\sum_{i} \hat{n}_{i\uparrow}\hat{n}_{i\downarrow}+\mu\sum_{i}
\hat{n}_{i}\,, \label{Hamiltonian}
\end{eqnarray}
where $t$ is the overall tunneling matrix element and $c_{i\sigma}$
($c^\dag_{i\sigma}$) denotes fermionic annihilation (creation)
operator for a fermion of spin $\sigma=\uparrow,\downarrow$ on the
lattice site $i$. The first term describes the nearest-neighboring
hoppings with the hopping matrices  given by
$\mathcal{R}_{ij}\equiv\exp[i{\bf A}\cdot({\bf r}_i-{\bf r}_j)]$,
where ${\bf A}=(\beta\sigma_y,\alpha\sigma_x,0)$ denotes a
non-Abelian gauge field which can be generated by the laser-induced
spin-flipped tunneling \cite{Osterloh,Ruseckas}. In this Letter we
set $\beta=-\alpha$, which implies that the SOC is of Rashba type
\cite{Grab,Cole,Radic,Cai,Gong,Zhang1,Qian,HLiang,Hickey}. In this
case, the spin-conserved hopping term is proportional to
$t\cos\alpha$, and the spin-flipped term is in proportion to
$t\sin\alpha$. $U$ is the on-site atomic repulsion and $\mu$ is the
chemical potential. The particle number operator is
$\hat{n}_{i}=\hat{n}_{i\uparrow}+\hat{n}_{i\downarrow}$ with
$\hat{n}_{i\sigma}=\hat{c}^\dag_{i\sigma}\hat{c}_{i\sigma}$.

{\it Method.}--We study the physical properties of Hamiltonian
(\ref{Hamiltonian}) with the non-perturbative cluster dynamical
mean-field theory (CDMFT), using Hirsch-Fye Quantum Monte Carlo
algorithm as the impurity solver   \cite{Hirsch1,Georges}. In the
presence of SOC, we can map the square lattice onto two sets of
sublattices for spin up (down) respectively, as shown in Fig.
\ref{fig2}(a). The $2\times2$ clusters are embedded in a
self-consistent medium with the  Weiss function of the cluster
represented by $g(i\omega)=\left(
            \begin{array}{cc}
             g_{\uparrow\uparrow} & g_{\uparrow\downarrow}\\
              g_{\downarrow\uparrow} & g_{\downarrow\downarrow}\\
            \end{array}
          \right)$,  where $g_{\sigma\sigma}$ and
          $g_{\sigma\bar{\sigma}}$ are the $4\times4$
matrix corresponding to spin conserved and spin flipped Weiss
functions. Due to the presence of the spin-flipping term in
Eq.~(\ref{Hamiltonian}), $g_{\uparrow\downarrow}$ and
$g_{\downarrow\uparrow}$ are generally nonzero. The CDMFT
incorporates spatial correlations and has been shown to be
successful in the study of MIT and   magnetic orders
\cite{Kotliar1,Maier,Kotliar2}. In this work, we shall investigate
the phase diagram on the half-filled square lattice with full range
of strength of atomic interactions and  SOC.

\begin{figure}[b]
\includegraphics[width=0.49\textwidth]{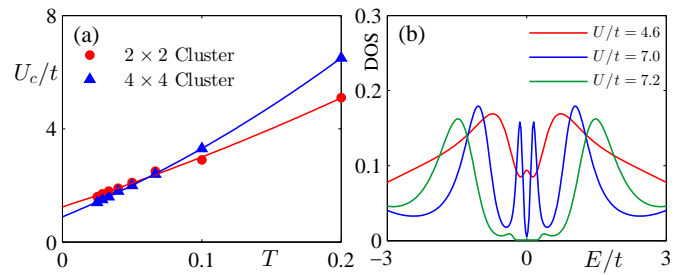}
\caption{(color online).  (a) Temperature scaling of the interaction
strength $U_c/t$ of the MIT for $\alpha=0$, with $2\times2$ and
$4\times4$ cluster repsecively. (b) Evolution of DOS at different
interaction strength $U/t$ for $\alpha=1.0$. } \label{fig3a}
\end{figure}

Before proceeding, we  first examine the case with $\alpha=0$ in
Hamiltonian (\ref{Hamiltonian}), which recovers the Hubbard model on
a conventional square lattice. At half filling, the Fermi surface
for non-interacting fermions is perfectly nested and the
antiferromagnetic (AF) fluctuations can drive the system into an
insulator with infinitesimal atomic interaction \cite{Hirsch}. In
the approach of CDMFT, it was demonstrated that   without AF
fluctuations, the MIT between the paramagnetic metal and the
paramagnetic Mott insulator occurs at $U_c/t\simeq6.05$ \cite{Park}.
Here in our simulations, we allow  magnetic orders to set in and
find that $U_c/t$ is greatly reduced as shown in Fig. \ref{phase1}.
Further, we perform a scaling analysis in Fig. \ref{fig3a}(a),
showing that the interaction strength $U_c/t$ at zero temperature
would approach to much smaller values in larger clusters.

{\it MIT.}--Now, we turn to the effects of the SOC on the MIT. We
concentrate on the basic region given by $\alpha\in[0,\pi/2]$ since
the relevant physical results are not affected in other regions.
First,  the single-particle spectrum is split into two bands [see
Fig. \ref{fig2}(b)], with the zero energy Fermi surface possessing a
particle and hole Fermi-pocket around the center and corner of the
Brillouin zone. The corresponding Density of states (DOS) for
non-interacting fermions is shown in Fig. \ref{fig2}(c), where the
zero energy DOS is suppressed and the bandwidth shrinks gradually
with increasing $\alpha$. The suppressed zero energy DOS reduce the
correlation effects on Fermi surface and hence enhance $U_c/t$ of
the MIT, whereas the shrinking bandwidth tends to make the  Mott
insulator happen at smaller $U_c/t$. The two effects compete with
each other, leading to the drastic changes of the MIT boundary in
the phase diagram. In Fig. \ref{phase1} we show that, away from
$\alpha=0$ the value of $U_c/t$ is rapidly increased due to the
suppression of the conventional AF fluctuation on square lattices.
Subsequently, the MIT exhibits a nonmonotonic behavior as a function
of $\alpha$. Specially at $\alpha=\pi/2$, the MIT occurs at a finite
atomic interaction with $U_c/t=4.1$.

In order to feature the MIT in the presence of SOC, Fig.
\ref{fig3a}(b) plots the evolution of DOS at different atomic
interactions for $\alpha=1.0$. We show that, compared to $\alpha=0$
case, the zero energy spectral peak in the metal phase (red line) is
largely suppressed by the SOC.  Simultaneously,  two satellite peaks
appear corresponding to the Van Hove singularity shown in Fig.
\ref{fig2}(c). Then, the zero energy peaks are gradually reduced and
a gap opens with the increase of atomic interactions.
\begin{figure}[b]
\includegraphics[width=0.47\textwidth]{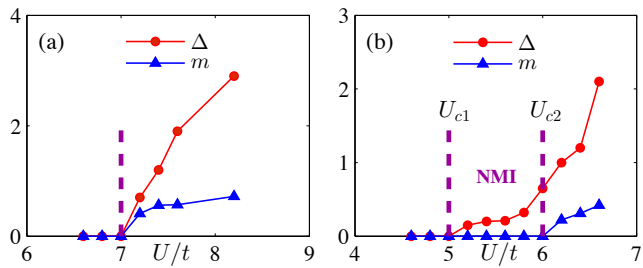}
\caption{(color online).   Single-particle gap $\Delta$ and
magnetization $m$ as functions of the interaction strength $U/t$ for
(a) $\alpha=1.0$ and (b) $\alpha=1.5$, respectively.} \label{fig3b}
\end{figure}

Fig. \ref{fig3b}(a) plots the corresponding single-particle gap
$\Delta$ and magnetization $m$ as functions of $U/t$ for
$\alpha=1.0$. The insulating phase characterized by a non-zero
$\Delta$ is accompanied by a finite $m$ simultaneously, indicating a
magnetic order arises. The specific magnetic phases in
Fig.~\ref{phase1} can be determined by identifying the spin
configurations on the cluster, see the Supplementary material for
more details. Fig. \ref{phase1} shows that, as $\alpha$ increases
the system transits from $xy$-antiferromagnet ($xy$-AFM) to spiral,
stripe, and spin vortex (SV) phases. Qualitatively, this can be
understood from an effective spin model \cite{Note}, where the
induced DM-type super-exchange term \cite{Dzaloshinsky,Moriya}
competes with the Heisenberg coupling, tending to form diverse spin
phases. However,  the effective spin model works only for the deep
Mott regime with the atomic kinetic energies being treated
perturbatively. In close proximity to the more interested Mott
transition, such a perturbative description breaks down and the
strong fluctuations arising from SOC may destroy the magnetic orders
and trigger an {\it order to disorder} transition. To address this
issue, one needs to implement a non-perturbative method such as
CDMFT to explore in detail the phase diagram as in Fig.
\ref{phase1}.

{\it NMI phase.}--To our surprise, despite the robustness of the
diverse spin phases in the Mott insulating regime with up to modest
values of SOC, a NMI phase is found to emerge in the vicinity of the
MIT for $\alpha>\alpha_c$ ($\alpha_c\simeq1.43$). The   NMI phase is
characterized in Fig. \ref{fig3b}(b), where the single-particle gap
$\Delta$ and magnetization $m$ occur for different atomic
interactions $U_{c_1}$ and $U_{c_2}$. Specifically in the
intermediate region $U_{c_1}\leq U\leq U_{c_2}$, the system enters
into an insulating state but with no long-range magnetic order. This
is a unique feature of the SO coupled square lattice for $\alpha$
being close to $\pi/2$, where the single-particle hopping becomes
nearly spin-flipped and the DOS is almost suppressed at zero energy.
\begin{figure}[t]
\includegraphics[width=0.4\textwidth]{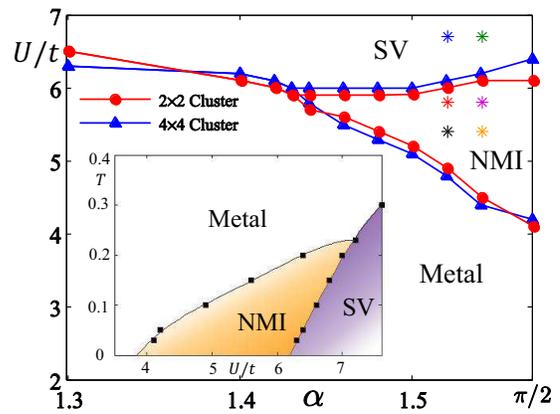}
\caption{(color online). Phase diagram in the vicinity of MIT for
$\alpha$ being close to $\pi/2$ and $T=0.05t$, obtained with
$2\times2$ and $4\times4$ cluster respectively. The stars mark
the parameters employed in Fig. \ref{fig4}. Inset: Temperature
dependence of the phase diagram for $\alpha=\pi/2$.  } \label{fig4a}
\end{figure}

The emergence of the NMI phase is further confirmed on a $4\times4$
cluster. The larger size of cluster incorporate more spin
correlations and thus, a better description of the atomic
correlations and SOC induced fluctuations can be expected. Fig.
\ref{fig4a} plots the  phase diagram for $\alpha$ being close to
$\pi/2$. We found that, in the $4\times4$ cluster, the regime of the
NMI phase is slightly expanded, demonstrating that the NMI is robust
in this system. We further show, in the inset of Fig. \ref{fig4a},
the temperature dependence of the NMI phase. The interval between
the metallic and SV phases enlarge with decreasing temperature. It
demonstrates that the NMI phase is more stable at low temperatures
by the suppression of thermal fluctuations.

The NMI phase breaks neither spin nor lattice symmetry, suggesting a
potential spin-liquid (SL) ground state.  Such a fundamental state
was first proposed by Anderson \cite{Anderson}   and has long been
sought in the frustrated spin systems \cite{Yan}. Recently,
interacting fermion models have attracted wide attentions
\cite{Meng,Lu,Clark,Yang}, and it was reported that a SL state can
be identified  on honeycomb lattice between semimetal and AF
insulator with $3.5t\leq U \leq 4.3t$ \cite{Meng}. Despite this, its
presence has been challenged since the interval of the SL phase is
small, which may vanishes under the size scaling
\cite{Sorella,Hassan,Assaad}. The latest results using large-scale
quantum Monte Carlo (QMC) showed that, if the SL state exists, the
possible   regime reduces substantially to a small interval
$3.8t\leq U \leq 3.9t$ \cite{Sorella}. Similar situations have been
encountered for the staggered-flux model on a square lattice
\cite{Chang,Otsuka}. Here, the essential feature characterizing the
present system is the considerably large space of parameters, where
the NMI phase emerges. This is in sharp contrast to   the limited
phase space ($3.4t\leq U \leq 3.9t$) obtained in the interacting
fermions on honeycomb lattices \cite{Wu}. Specially, the predicted
NMI phase occurs until $\alpha>\alpha_c$, showing that it is a
strong field effect of the SOC.
\begin{figure}[t]
\includegraphics[width=0.4\textwidth]{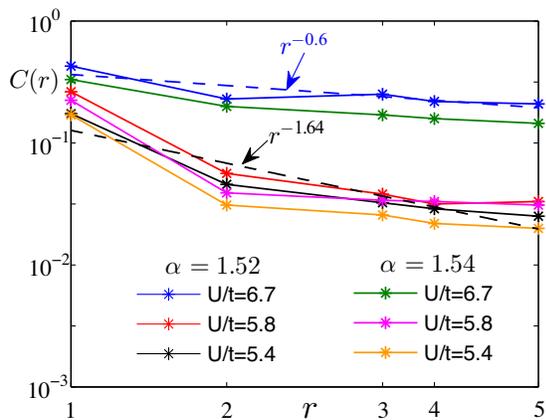}
\caption{(color online).  Staggered spin-spin correlation function
for $\alpha=1.52$ and  $1.54$ with different interaction strength
 $U/t$ marked by the stars in Fig. \ref{fig4a}. The dashed lines
are representative power-law fits to the data. } \label{fig4}
\end{figure}

The absence of magnetic orders in the NMI phase implies strong
short-range spin correlation. However, it may decay as a power-law
or exponentially. To explore this issue, we calculate the staggered
spin-spin correlation function
\begin{eqnarray}
C(\mathbf{r})=(-1)^{\mathbf{r}}\langle {\rm S}^x_0 {\rm
S}^x_{\mathbf{r}}+{\rm S}^y_0 {\rm S}^y_{\mathbf{r}}+{\rm S}^z_0
{\rm S}^z_{\mathbf{r}}\rangle\,. \label{spin-correlation}
\end{eqnarray}
as shown in Fig. \ref{fig4}, where the  spin correlation functions
are fitted to a power-law as $C(\mathbf{r})\sim1/r^\gamma$. In NMI
phase, we find that the exponent $\alpha$ is less than $2$ with
$\alpha\sim1.6$ in our simulations. Whereas in the deep Mott
insulating regime   where spin is ordered, $\alpha$ becomes much
smaller. Therefore, the NMI phase seems to suggest a candidate of
algebraic SL. Further studies would be implemented by QMC
calculations in the future.

The above  phenomena of  the intriguing MIT and exotic matter states
can be investigated in experiments. In  optical lattices, the Mott
insulating phase can be detected by site-resolved imaging of single
atoms \cite{Bakr2,Gemelke,Bakr1,Sherson,Bloch1}, and the spin
textures occurring in the Mott pahse can be observed via {\it in
situ} microscopy \cite{Weitenberg} or through spin-resolved
time-of-flight measurements \cite{Lin2}.  On the other hand, the
spin correlation can be measured by the spin structure factors in
optical Bragg scattering \cite{Mathy},  which may present the
signatures of the spin ordered phases and the power-law scaling of
the NMI phase.  In addition, an extremely low-temperature  has been
recently realized to approach the superexchange energy scales
\cite{Murmann}.

In summary, we have investigated the half-filled  Fermi Hubbard
model with Rashba-type SOC on a square lattice. We show that this
system displays a rich phase diagram.  The interplay between the
atomic interactions and  SOC results in distinctive features of the
MIT with diverse spin ordered phases occurring on the side of Mott
insulator. Near the Mott transition, a quantum NMI phase is found to
emerge in a considerably large regime of parameters due to the
strong field effect of the SOC, formulating a new avenue to achieve
the intriguing quantum disordered state beyond the spin systems.

\begin{acknowledgments}

We would like to thank G. Juzeli\={u}nas, X. C. Xie, N. H. Tong,  X.
S. Yang, and X. F. Zhang for many helpful discussions. This work is
supported by NCET, NSFC under grants Nos. 11474205, 11404225. We
acknowledge the supercomputing center of CAS for the computational
resources.
\end{acknowledgments}


\begin{thebibliography}{99}
\bibitem{Note}
For $U/t\gg 1$,  we can apply the second order perturbation theory
to the system and obtain
$\hat{H}_{\textnormal{eff}}=\sum_{i,\delta=\hat{x},
\hat{y}}[\sum_{a=x,y,z}J_{\delta}^{a}\hat{S}_{i}^{a}
\hat{S}_{i+\delta}^{a}+\mathbf{D_{\delta}}\cdot(\hat{\mathbf{S}}_{i}
\times\hat{\mathbf{S}}_{i+\delta})]$, with $J_{\hat{x}}^{x,z}=
J_{\hat{y}}^{y,z}=4t^{2}\cos(2\alpha)/U$,
$J_{\hat{x}}^{y}=$$J_{\hat{y}}^{x}=4t^{2}/U$,
$\mathbf{D}_{\hat{x}}=4t^{2}\sin(2\alpha)\hat{y}/U$, and
$\mathbf{D}_{\hat{y}}=4t^{2}\sin(2\alpha)\hat{x}/U$. Here, the first
term is the conventional Heisenberg coupling and the second term
denotes the so-called Dzyaloshinskii-Moriya (DM)-type super-exchange
\cite{Dzaloshinsky,Moriya}.
\end{thebibliography}
\end{document}


\title{Supplementary Material\\
Emergence of Quantum Nonmagnetic Insulating Phase in Spin-Orbit
Coupled Square Lattices}
\author{Xin Zhang}
\affiliation{Center for Theoretical Physics, Department of Physics,
Capital Normal University, Beijing 100048, China}
\author{Wei Wu}
\affiliation{D\'{e}partement de Physique and RQMP, Universit\'{e} de
Sherbrooke, Sherbrooke, Qu\'{e}bec, Canada}
\author{Gang Li}
\affiliation{Lehrstuhl fuer Theoretische Physick, Universit\"{a}t
Wuerzburg, 97074 Wuerzburg, Germany }
\author{Lin Wen}
\affiliation{College of Physics and Electronic Engineering,
Chongqing Normal University, Chongqing, 401331, China}
\author{Qing Sun}
\affiliation{Center for Theoretical Physics, Department of Physics,
Capital Normal University, Beijing 100048, China}
\author{An-Chun Ji}
\affiliation{Center for Theoretical Physics, Department of Physics,
Capital Normal University, Beijing 100048, China}
\date{{\small \today}}

\maketitle

In this Supplementary Material, we describe how to determine the
diverse spin phases and the corresponding magnetization. First, in
the approach of CDMFT,  the Weiss function of the $N$-site cluster
embedded in a self-consistent medium is  determined by the cluster
self-energy $\Sigma(i\omega)$  via the coarse-grained Dyson equation
\cite{Maier,Kotliar2}
\begin{eqnarray}
g^{-1}(i\omega)=\left[\sum_{\bf K}\frac{1}{i\omega+\mu-t({\bf
K})-\Sigma(i\omega)}\right]^{-1}+\Sigma(i\omega)\,, \label{Dyson}
\end{eqnarray}
where  $t({\bf K})$ is the Fourier-transformed  hopping matrix with
wave vector ${\bf K}$ in the cluster reduced Brillouin zone of the
superlattice. Note that, in the presence of SOC both the above Weiss
function and the  self-energy of the cluster in Eq. (\ref{Dyson})
become  $g(i\omega)=\left(
            \begin{array}{cc}
             g_{\uparrow\uparrow} & g_{\uparrow\downarrow}\\
              g_{\downarrow\uparrow} & g_{\downarrow\downarrow}\\
            \end{array}
          \right)$ and $\Sigma(i\omega)=\left(
            \begin{array}{cc}
             \Sigma_{\uparrow\uparrow} & \Sigma_{\uparrow\downarrow}\\
              \Sigma_{\downarrow\uparrow} & \Sigma_{\downarrow\downarrow}\\
            \end{array}
          \right)$.
In this case, we  introduce two-component  row and column fermionic
field operators:
$\hat{\Psi}^{\dag}=[\hat{c}_{i\uparrow}^\dag,\hat{c}_{i\downarrow}^\dag]$
and $\hat{\Psi}=[\hat{c}_{i\uparrow},\hat{c}_{i\downarrow}]^T$, and
define the cluster Green's function  as
\begin{eqnarray}
G_{\sigma\sigma^\prime}(\tau)=\langle\Psi(\tau)\Psi^\dag(0)\rangle=\left(\begin{array}{cc}
             G_{\uparrow\uparrow}(\tau) & G_{\uparrow\downarrow}(\tau)\\
              G_{\downarrow\uparrow}(\tau) & G_{\downarrow\downarrow}(\tau)\\
            \end{array}\right)\,.
\label{spinor}
\end{eqnarray}
Once $g(i\omega)$ is determined, the impurity solver can be used to
compute the cluster Green's function $G(i\omega)$. Eventually, by
using the Dyson equation
$\Sigma(i\omega)=g^{-1}(i\omega)-G^{-1}(i\omega)$, the
self-consistent iterative  $G(i\omega)$ is  obtained.

The energy gap $\Delta$ can be derived by the  local density of
states (LDOS).  By implementing the analytic extension of the
imaginary time cluster Green's function  $G(i\omega)$ via the
maximum entropy method \cite{Jarrell}, we have
\begin{eqnarray}
\rho(\omega)=\sum_{\bf k} A({\bf k},\omega)\approx-\frac{1}{\pi}{\rm
Im}[G_{ii}(\omega)]\,.
\end{eqnarray}
Then in the spectrum of LDOS, we can obtain the  energy gap $\Delta$
by the energy  width of zero density of states.

The different spin phases  in the Mott insulating regime can be
characterized by the  spin structure factor $S_{\bf{q}}=|\sum_i{\bf
S}_ie^{i{\bf q}\cdot {\bf r}_i}|$ with ${\bf{q}}$ the  2D wave
vector. Here, ${\bf S}_i=\langle \hat{\bf S}_i\rangle$ denotes local
magnetic order parameter on site $i$ of the cluster, with three
components given by
\begin{eqnarray}
&&S_i^x=\frac{1}{2}\langle c_{i\uparrow}^\dag c_{i\downarrow}+
c_{i\downarrow}^\dag
c_{i\uparrow}\rangle=\frac{1}{2}{\rm Re}
[G_{i,\uparrow\downarrow}(0^+) +G_{i,\downarrow\uparrow}(0^+)]\,,\\
&&S_i^y=-\frac{1}{2}i\langle c_{i\uparrow}^\dag c_{i\downarrow}-
c_{i\downarrow}^\dag c_{i\uparrow}\rangle=-\frac{1}{2}{\rm
Im}[G_{i,\uparrow\downarrow}(0^+)
-G_{i,\downarrow\uparrow}(0^+)]\,,\\
&&S_i^z=\frac{1}{2}\langle c_{i\uparrow}^\dag c_{i\uparrow}-
c_{i\downarrow}^\dag c_{i\downarrow}\rangle=\frac{1}{2}{\rm
Re}[G_{i,\uparrow\uparrow}(0^+) -
G_{i,\downarrow\downarrow}(0^+)]\,.
\end{eqnarray}
The  spin ordered phases in Fig. 1 of the main text are derived on
$2\times2$ cluster, where the structure factor of the $xy$-AFM has a
peak at ${\bf{q}}=(\pi,\pi)$, the stripe phase at
${\bf{q}}=(0,\pi)$, and the SV phase at ${\bf{q}}=(\pi,0)$ and
${\bf{q}}=(0,\pi)$. Between the $xy$-AFM and stripe phases, spiral
phases where the spins spiral in the $z$-${\bf{q}}$ plane with
${\bf{q}}=(q,\pi)$ the in-plane wave vector may appear. However, the
spiral phase is hard to be explicitly identified on $2\times2$
cluster. To overcome this difficulty, we explore on a larger
$4\times4$ cluster, and a spiral-4 phase with spatial period of
$4\times2$ lattice sites is clearly identified in the following FIG.
1.

Finally, we present the definition of magnetization. In the  spin
ordered phases, we can rotate the local magnetic order parameter
${\bf S}_i$ on each cluster site to a global coordinate system:
${\bf S}_i^\prime=U(\phi_i){\bf S}_i$, with $\phi_i$ the angle
between the local and global coordinates. Then we can define ${\bf
m}=\frac{1}{N}\sum_{i=1}^N{\bf S}^\prime_i$ with the magnetization
given by $m^2=\sum_{a=x,y,z} m_a^2$. For example, in the $xy$-AFM
shown in Fig. 1, there are two sublattices $(\phi_{i\in
A}=0,\phi_{i\in B}=\pi)$, we have ${\bf
m}=\frac{1}{N}\sum_{i=1}^N\epsilon_i{\bf S}_i$, where
$\epsilon_i=\pm1$ is for sites belonging to sublattice $A(B)$
respectively. This general definition of magnetization is also
applied to other spin  phases throughout this paper.
\begin{figure}[h]
\includegraphics[width=0.46\textwidth]{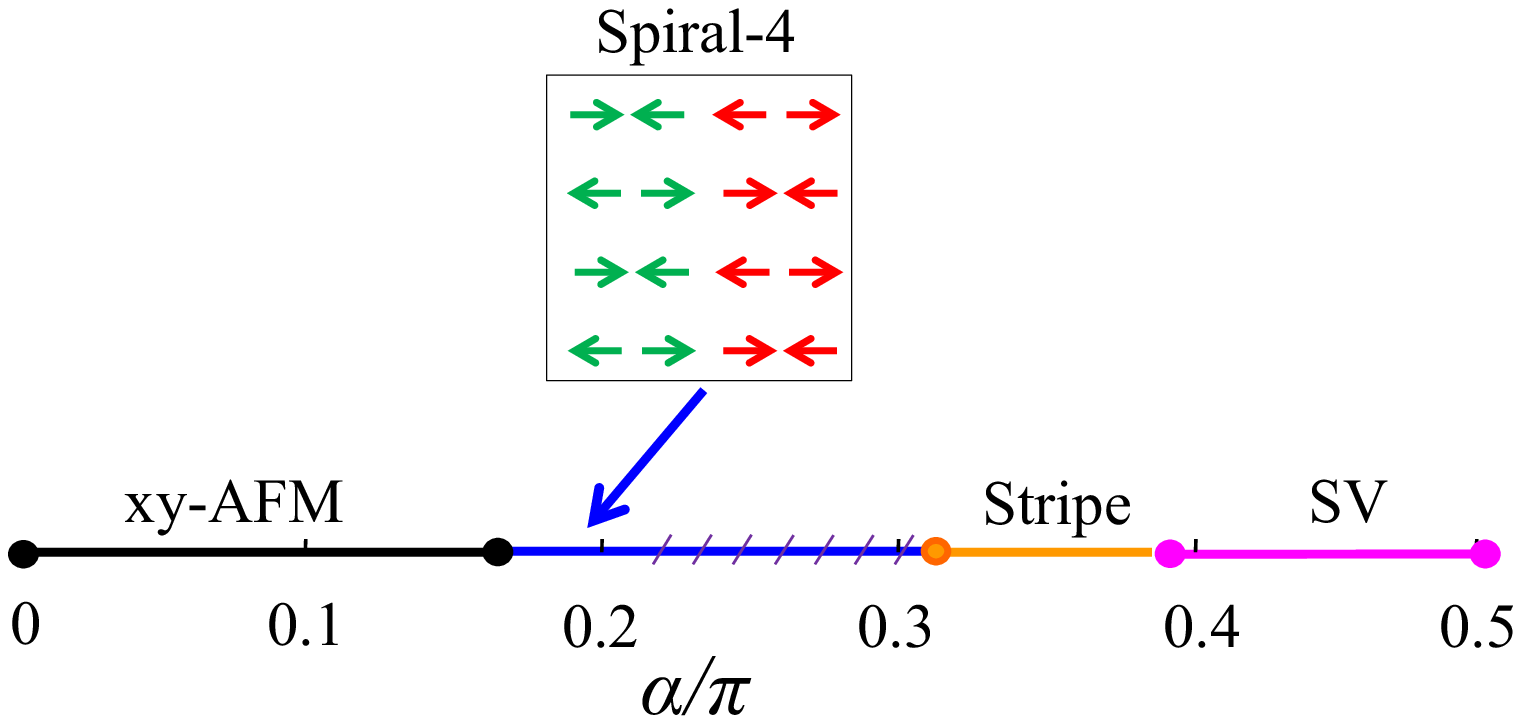}%
\caption{(color online). Spin phase diagram in the Mott insulating
regime with  $U/t=7.5$ on the $4\times4$ cluster. The intervals of
the $xy$-AFM, SV and stripe phases agree well with those on
$2\times2$ cluster. Specifically, a spiral-4 phase  with spatial
period of $4\times2$ lattice sites (the green and red arrows
indicate the spins have up or down $z$-components) is explicitly
identified between the  $xy$-AFM and stripe phases. The shaded area
indicate other commensurate or non-commensurate spiral phases.
\label{ffs}}
\end{figure}